\documentclass[aps,prl,twocolumn]{revtex4-1}

\usepackage{graphics}
\usepackage{epsfig}
\usepackage{bm}
\usepackage{color}

\begin{document}

%Include all the eps files in Figures directory
\graphicspath{{Figures/}}

\title{Time-reversal anomaly and Josephson effect in time-reversal invariant topological superconductors}

\author{Suk Bum Chung}
\author{Joshua Horowitz}
\author{Xiao-Liang Qi}
\affiliation{$^1$Department of Physics, Stanford University, Stanford, CA 94305, USA}

\date{\today}

\begin{abstract}
Topological superconductors are gapped superconductors with protected Majorana surface/edge states on the boundary. In this paper, we study the Josephson coupling between time-reversal invariant topological superconductors and $s$-wave superconductors. The Majorana edge/surface states of time-reversal invariant topological superconductors in all physical dimensions 1, 2, 3 have a generic topological property which we named as time-reversal anomaly. Due to the time-reversal anomaly, the Josephson coupling prefers a nonzero phase difference between topological and trivial superconductors. The nontrivial Josesphon coupling leads to a current-flux relation with a half period in a SQUID geometry, and also a half period Fraunhofer effect in dimension higher than one. We also show that an in-plane magnetic field restores the ordinary Josephson coupling, as a sharp signature that the proposed effect is a consequence of the unique time-reversal property of the topological edge/surface states. Our proposal provides a general approach to
%We also predicted other experimental signatures including the Fraunhofer effect and the effect of in-plane magnetic field, which can %be
experimentally verify whether a superconductor is topological or not. %by applying current experimental techniques to topological superconductors.
\end{abstract}

%\pacs{05.10.Cc,75.10.Jm,71.10.-w}

\maketitle

%\section{Introduction}

In recent years, topological states of matter such as topological insulators (TI) and topological superconductors (TSC) have attracted tremendous theoretical and experimental interest\cite{qi2011rmp,hasan2010,moore2010}. %A topological states of matter is defined as a zero-temperature state that cannot be adiabatically deformed to a trivial state without experiencing a phase transition, even if it has the same symmetry as the trivial state.
%TSC's are fully gapped superconductors with a bulk gap for the quasi-particle excitations, but robustly gapless surface/edge states protected by bulk topology. The surface/edge states are Majorana fermions, meaning that the surface quasiparticle is its own anti-particle.
The first example of TSC is the time-reversal breaking $(p+ip)$-wave superconductor of spinless fermions in two-dimensions\cite{read2000}. More recently, a new class of TSC was proposed in three-dimensional (3d) time-reversal invariant (TRI) superconductors\cite{schnyder2008,roy2008,qi2009b}, which has time-reversal invariant pairing order parameter in the bulk, and two-dimensional massless Majorana fermions with linear dispersion on the boundary. Candidate materials of TSC include the $^3$He B phase\cite{schnyder2008,roy2008,qi2009b,chung2009}, ${\rm Cu}$-doped ${\rm Bi_2Se_3}$\cite{fu2010b,hsieh2012,sasaki2011} and $p$-type ${\rm TlBiTe_2}$\cite{yan2010a}.

%The $^3$He B phase was proposed as an example of the 3d TRI TSC (more rigorously, a topological superfluid)\cite{schnyder2008,roy2008,qi2009b,chung2009}. Solid state candidates of 3d TSC have also been proposed in doped topological insulators including ${\rm Cu}$-doped ${\rm Bi_2Se_3}$\cite{fu2010b} and $p$-type ${\rm TlBiTe_2}$\cite{yan2010a}. In a recent tunneling experiment\cite{sasaki2011}, zero bias peak is observed which is consistent with the theoretical proposal\cite{hsieh2012}, although more experiments are needed to reach a conclusion. %XLnote: Can be shorten further.

A robust topological state of matter should be characterized by topological effects that qualitatively distinguishes it from trivial states. The first topological effect for TRI TSC was proposed in Ref. \cite{qi2009b} which considered the time-reversal property of a pair of Majorana zero modes in a TRI vortex defect of 2D TSC. As a consequence of the Majorana zero modes, the time-reversal symmetry maps a state with odd number of fermions in the zero modes to one with even number of fermions. This is due to the following equation
 \begin{equation}
\mathcal{T}^{-1} i\gamma_{0\uparrow}\gamma_{0\downarrow} \mathcal{T} = -i\gamma_{0\uparrow}\gamma_{0\downarrow}
\label{EQ:SUSY}
\end{equation}
where $\gamma_{0\uparrow,\downarrow}$ are the Majorana zero modes which form a Kramers pair: $\mathcal{T}^{-1}\gamma_{0\uparrow}\mathcal{T}=\gamma_{0\downarrow}, \mathcal{T}^{-1}\gamma_{0\downarrow}\mathcal{T}=-\gamma_{0\uparrow}$. $i\gamma_{0\uparrow}\gamma_{0\downarrow}$ is the fermion number parity operator with eigenvalue $\pm 1$. The fact that time-reversal transformation changes fermion number parity is generic for any Kramers pair of Majorana zero modes, which applies to the edge/surface states of TRI TSC in all physical dimensions $D=1,2,3$. Since fermion number should be preserved microscopically by time-reversal transformation, the anti-commutation between time-reversal and fermion number parity given by Eq. (\ref{EQ:SUSY})  is an emergent property and cannot be modified by any local perturbation to the TSC. This property requires the gaplessness of the surface states, since the lowest energy state with even and odd number of fermions are required to be degenerate. Conversely, opening a gap at the surface state necessarily breaks TR symmetry. We name this property of TSC as {\it time-reversal anomaly} due to its similarity to the parity anomaly\cite{niemi1983, redlich1984}. The time-reversal anomaly can in principle be measured if one can measures the fermion number parity near an edge. However, it is practically difficult to measure fermion number parity. Other topological effects have been proposed\cite{wang2011,ryu2012b,qi2012b} % in topological field theory approaches, including gravitational topological response\cite{wang2011,ryu2012b} and axionic coupling between superconducting phase order parameter and the electromagnetic field\cite{qi2012b},
but their experimental observation are also highly nontrivial.

%Topological effects have been proposed in time-reversal TSC, including topological response to gravitational field\cite{wang2011,ryu2012b} and one-dimensional chiral Majorana fermions propagating along domain walls between different surface T-breaking regions\cite{teo2009,wang2011,qi2012b}. In particular, Ref. \cite{wang2011} proposed the setting of two $s$-wave superconductor films deposited on top of a TSC. When the two $s$-wave superconductors have the phase difference of $\theta$ and $-\theta$ respect to the TSC, time-reversal symmetry is broken on the surface and the number of chiral Majorana zero modes on the domain wall between the two regions is uniquely determined by the bulk topological invariant of the TSC. Such a topological effect can be characterized by a topological field theory\cite{qi2012b}, in which the phase of the superconducting order parameter on each fermi surface of the TSC is coupled to electromagnetic field as an axion\cite{peccei1977}. Due to its charge neutrality, the chiral Majorana fermion predicted in this setting can only be measured through thermal transport.

In this paper, we translate the time-reversal anomaly into a topological Josephson effect
%propose a new topological effect in TRI TSC
which is easy to measure in current experimental techniques, and can be used to distinguish TSC from ordinary SC in all physical spatial dimensions $D=1,2,3$. Similar to Refs. \cite{wang2011,qi2012b}, we consider the Josephson coupling between TSC and $s$-wave superconductors. The regular Josephson coupling in such a junction is very weak due to different pairing symmetry, but
%Naively one would have expected that the Josephson coupling between TSC and $s$-wave superconductor almost vanishes due to different pairing symmetries. However
the Majorana surface states induce a nontrivial Josephson coupling which is qualitatively different from the ordinary case. Since the surface state is protected by time-reversal symmetry, it remains gapless if the relative phase between TSC and $s$-wave superconductor is $0$ or $\pi$, while it generically becomes gapped for $\theta\neq 0,\pi$. Since opening a gap in the surface states can save energy, this simple reasoning leads to the dramatic conclusion that the lowest energy state of such Josephson junction does not occur at relative phase $\theta=0$ or $\pi$, but somewhere between $0$ and $\pi$. Furthermore, the time-reversal anomaly Eq. (\ref{EQ:SUSY}) tells us that the ground state at relative phase $\theta$ and $-\theta$ have opposite fermion number parity, since they are related by time-reversal. Consequently the system cannot stay at the lowest energy state if the fermion number at the junction is conserved. We studied the behavior of the topological Josephson coupling for both systems with a conserved fermion number and those with a fluctuating fermion number which may be induced by disorder states in a real system. For fixed fermion number the Josephson effect has a phase shift determined by the fermion number. For fluctuating fermion number the Josephson effect has a doubled frequency. In both cases the junction is qualitatively different from an ordinary Josephson junction, which can be detected in the behavior of a SQUID (superconducting quantum interference device) and in the Fraunhofer effect (for %a single junction with a magnetic field
$D=2,3$). In particular, in $D=1$ or $D=2$ we show that an in-plane magnetic field comparable with the pairing gap at the Josephson junction can drive the junction back to a normal one, which provides a clear signature that time-reversal symmetry plays an essential role in the topological Josephson effect.\footnote{It should be clarified that the topological Josephson effect discussed in this work is qualitatively different from the fractional Josephson effect discussed in Ref. \cite{fu2008} which applies to a different physical system and has a halved rather than doubled frequency.}

\noindent{\bf  Model Hamiltonians}

Although our proposal applies to a generic TSC, for the concreteness of our discussion we would like to start by defining some model systems for TRI TSC. The TSC in $D=1, 2, 3$ can all be realized in the following
%One convenient way to formulate a lattice model of TRI $p$-wave TSC in 1, 2, 3 dimensions is to start with the 3D cubic lattice model of $^3$He-B model. The
lattice Bogoliubov-de Genne (BdG) Hamiltonian
\begin{eqnarray}
H_{p} =&&-\mu \sum_{{\bf r}s}c^\dagger_{{\bf r}s}c_{{\bf r}s} - t \sum_{i=1}^D\sum_{{\bf r}s}(c^\dagger_{{\bf r}+{\bf \hat{e}}_i s}c_{{\bf r}s} + {\rm h.c.})\nonumber\\
&& + \sum_{i=1}^D\sum_{{\bf r}ss'}[\Delta (-{\bf \hat{e}}_i \cdot {\bm \sigma} \sigma_2)_{ss'} c^\dagger_{{\bf r}+{\bf \hat{e}}_i s}c^\dagger_{{\bf r}s'} + {\rm h.c.}],
\label{EQ:BWd}
\end{eqnarray}
In $D=2$ (3), the model is defined on a square (cubic) lattice, so
%Here %$D=1,2,3$ denotes the spatial dimension.
${\bf r}$ denotes the lattice sites and $\hat{\bf e}_i$'s ($i=1,2,..,D$) are the orthogonal unit vectors. ${\bm \sigma}=\left(\sigma_x,\sigma_y,\sigma_z\right)$ denotes the Pauli matrices, and $s=\uparrow,\downarrow$ the electron spin index. The TRI TSC phase requires the chemical potential is in the range $-2Dt <\mu<-2(D-2)t$. %However, the topological invariant is dimension dependent: $N=1$ for $D=3$ and $Z_2$ nontrivial for $D=1,2$.

%XLnote ({\bf 1. depending on the later part, decide whether we need the rest of this section. 2. Consider to add a figure on band structure and surface states.})

\noindent{\bf Josephson effect in 1D TSC}

We start from 1D TSC, which gives us the clearest example of how the Josephson coupling between the TRI TSC and the $s$-wave SC arises from time-reversal symmetry breaking. 1D is distinct from 2D and 3D in the sense that there is a finite gap separating the Majorana zero modes with other quasi-particle excitations. The candidate material for the 1D TRI TSC include a class of quasi-1D organic superconductors \cite{lee1997} and LiMO \cite{mercure2012}. %({\bf Ref on LiMO?}) 
As will be discussed below, such a gap makes it possible to probe the ``TR anomaly" of Eq.(\ref{EQ:SUSY}), {\it i.e.}, the fact that time-reversal changes the fermion number parity of the Majorana zero modes. %XLnote: consider to introduce the concept of TR anomaly in introduction.

The 1D Josephson junction we consider here is described by the Hamiltonian $H=H_p+H_s+H_1$, with $H_p$ the Hamiltonian of a 1D TRI TSC chain given by Eq.\ref{EQ:1Dtsc}, $H_s$ that of an s-wave superconducting chain
%\begin{eqnarray}
%H_s =&-&\mu' \sum_{n\sigma} f^\dagger_{n\sigma} f_{n\sigma} - t'\sum_{n\sigma} (f^\dagger_{n+1,\sigma} f_{n,\sigma}+ {\rm h.c.})\nonumber\\
%&+&  \sum_n (\Delta' f^\dagger_{n\uparrow} f^\dagger_{n\downarrow} + {\rm h.c.}).
%\end{eqnarray}
and $H_1$ a weak spin-conserving hopping between the first site of TSC chain and the last site of $s$-wave chain:
\begin{equation}
H_1 = - \sum_\sigma[(\delta t)c^\dagger_{1\sigma}f_{m\sigma} + {\rm h.c.}].
\label{EQ:interHop}
\end{equation}
%({\bf please fix the notation for the last site. Somewhere it is $m$ somewhere it is $N$. I suggest to use $N$.})
%that conserves the spin. %For the simplicity of calculation, we will

%\subsection{Perturbation theory}

%{\bf Fig 1: (a) Schematic picture of the SQUID. (b) $I-\phi$ relation. (c) $I_c-\Phi$ relation of SQUID (for fixed and fluctuating fermion number parity). (d) Cross-over induced by magnetic field.}

For small coupling $\delta t$ we can study the Josephson coupling by perturbation theory. Since the TSC and the $s$-wave SC are both gapped, the main effect of a small $\delta t$ is to couple the Majorana zero modes at the boundary of TSC. We start from a special case $\mu = 0, \Delta = t$ for the p-wave chain, in which case the TSC Hamiltonian can be written as
%\begin{equation}
$H_p = -t\sum_{n\sigma}(c_{n+1,\sigma}+c^\dagger_{n+1,\sigma})(c_{n,\sigma}-c^\dagger_{n,\sigma})$ \cite{kitaev2001},
%\label{EQ:localBdG}
%\end{equation}
in which case the Majorana zero modes are completely localized at the boundary site: $\gamma_{0\sigma} = c_{1\sigma}+c^\dagger_{1\sigma}$. Therefore the interchain hopping $H_1$ can be written as
%\begin{equation}
$H_1 = -\sum_\sigma [\frac{\delta t}{2}\{\gamma_{0\sigma}-(c_{1,\sigma}-c^\dagger_{1,\sigma})\}f_{m\sigma} + {\rm h.c.}]$. %-\frac{\delta t}{2} \sum_\sigma [\gamma_{0\sigma}(f_{m\sigma}-f^\dagger_{m\sigma}) -(c_{1,\sigma}-c^\dagger_{1,\sigma})(f_{m\sigma}+f^\dagger_{m\sigma})].
%\label{Htzeromode}%+ \eta \sum_\sigma (c'^\dagger_{1,\sigma}c_{N,\bar{\sigma}}+{\rm h.c.}),
%\end{equation}
%and only the first term acts on the zero mode $\gamma_{0\sigma}$.
Only %keep
the zero mode terms contributes to the second order perturbation, %theory
giving the simple effective Hamiltonian
%By considering only the time-reversal symmetry, we can deduce that the effective Hamiltonian for this 1D Josephson junction is in the form
\begin{eqnarray}
\label{EQ:M1Josephson}
H_{eff} &=& J (i\gamma_{0\uparrow}\gamma_{0\downarrow}) \sin \phi.\\
iJ %&=&\frac{\delta t^2}{2}\int dt\left\langle T \left(f_{m\uparrow}-f_{m\uparrow}^\dagger\right)(t)\right.\nonumber\\
%& &\left.\cdot\left(f_{m\downarrow}-f_{m\downarrow}^\dagger\right)(0)\right\rangle_{\Delta'=i\left|\Delta'\right|}\nonumber\\
&=&(\delta t)^2\int dt\left\langle T f_{m\uparrow}(t)f_{m\downarrow}(0)\right\rangle_{\Delta'=i\left|\Delta'\right|},\nonumber
\end{eqnarray}
%({\bf SukBum please check this equation})
where $\Delta'$ is the pairing gap of the $s$-wave chain and $\phi$ its phase; %(Notice that
our gauge is set to give us a real $\delta t$ and $\langle T \ldots \rangle$ denotes the time-ordered expectation value. %The form of
This effective Hamiltonian has time-reversal symmetry, both $\sin\phi$ and $i\gamma_{0\uparrow}\gamma_{0\downarrow}$ being time-reversal odd.

The effective Hamiltonian of Eq.(\ref{EQ:M1Josephson}) is valid for generic parameters $\mu,\Delta$, for which the Majorana zero mode $\gamma_{0\sigma}$ is not completely localized at the first site. This is because the zero mode term of the inter-chain coupling $H_1=-u_0\frac{\delta t}{2}\sum_\sigma (\gamma_{0\sigma}f_{m\sigma}+ {\rm h.c.})+...$ remain identical up to a constant factor that comes from the mode expansion $c_{1\sigma}=u_0\gamma_{0\sigma}+\sum_{n\neq 0}(u_n\gamma_{n\sigma}+v^*_n \gamma^\dagger_{n\sigma})$, where $\gamma_{n\sigma} (\gamma^\dagger_{n\sigma})$'s are annihilation (creation) operators of finite energy quasi-particles.  %with $...$ denotes the terms involving only finite energy modes. Compare with Eq. (\ref{Htzeromode}) we see that the effective Hamiltonian of the zero modes has the same form as Eq. (\ref{EQ:M1Josephson}), and the only difference is $\delta t^2/4$ is replaced by $u_0^2\delta t^2$ in the value of $J$. %We also studied the generic case by exact diagonalization ({\bf cite figures}), which also agrees qualitatively with Eq. (\ref{EQ:M1Josephson}).

%One can immediately see that, due
Due to its dependence on the local fermion number parity $i\gamma_{0\uparrow}\gamma_{0\downarrow}$, the effective Hamiltonian Eq.(\ref{EQ:M1Josephson}) can give us %an unconventional Josephson coupling. %is dramatically different from that of an ordinary junction.
a nonzero Josephson current at $\phi=0,\pi$. This is because there is a limit where $i\gamma_{0\uparrow}\gamma_{0\downarrow}$ is quasi-conserved so %its eigenstates, with $i\gamma_{0\uparrow}\gamma_{0\downarrow}=\pm 1$, have energies of
our Josephson coupling should be taken to be $E_\pm=\pm J\sin\phi$ for $i\gamma_{0\uparrow}\gamma_{0\downarrow}=\pm 1$ respectively. This gives us the Josephson current of  $I(\phi)=\pm I_c\cos\phi$, where $I_c \equiv 2e J/\hbar$, which is nonzero at $\phi=0,\pi$. This is the physical signature of our TR anomaly - the local fermion number parity eigenstate breaking time-reversal symmetry. We emphasize that this is %not a special property of the model we use, but
a generic property of the TRI TSC. %since for $\phi=0,\pi$ the degeneracy between the two states of $i\gamma_{0\uparrow}\gamma_{0\downarrow}=\pm 1$ is protected by TR symmetry.

The above Josephson current from TR anomaly is physically observable for a sufficiently large Josephson frequency $\omega_J=2eV/\hbar$. %Now, $i\gamma_{0\uparrow}\gamma_{0\downarrow}$
In a clean system, the local fermion number parity is absolutely conserved at $T=0$, while in a realistic system, it may change, {\it e.g.} by a fermion leaking out of the zero modes to some impurity sites. We denote the rate of fermion parity changing process as $1/\tau$. Therefore in the fast limit $\omega_J\gg 1/\tau$, $i\gamma_{0\uparrow}\gamma_{0\downarrow}$ would not change in many Josephson period, {\it i.e.} it is quasi-conserved. On the other hand, in the slow limit $\omega_J\ll 1/\tau$, the system would stay at the ground state energy $E_G=-J\left|\sin\phi\right|$ of the the effective Hamiltonian Eq.(\ref{EQ:M1Josephson}), as it has enough time to relax into the lowest energy state. This gives us the current phase relation of $I(\phi)=I_c\frac{d}{d\phi}\left(-\left|\sin\phi\right|\right)=-I_c\frac{|\sin\phi|}{\sin\phi}\cos\phi$, which has doubled frequency $2\omega_J$ with discontinuities at $\phi=0,\pi$.

%XLnote: I moved the fast/slow discussion to the single junction part which makes it simpler.

\begin{figure}
\includegraphics[width=1.7in]{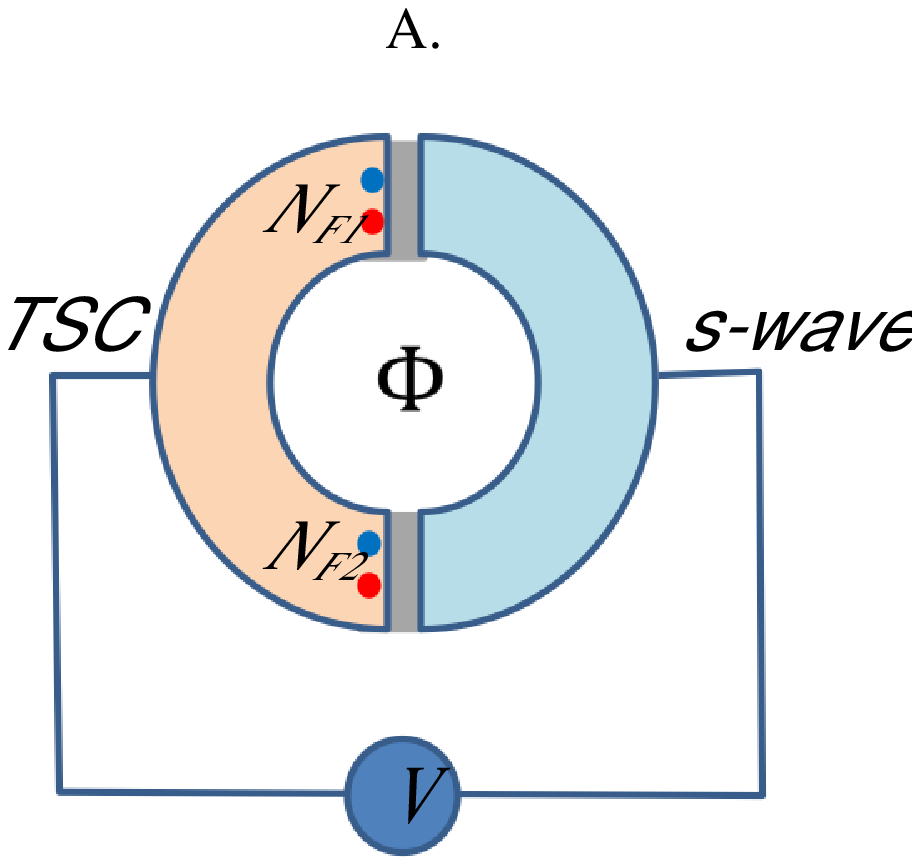}\includegraphics[width=1.7in]{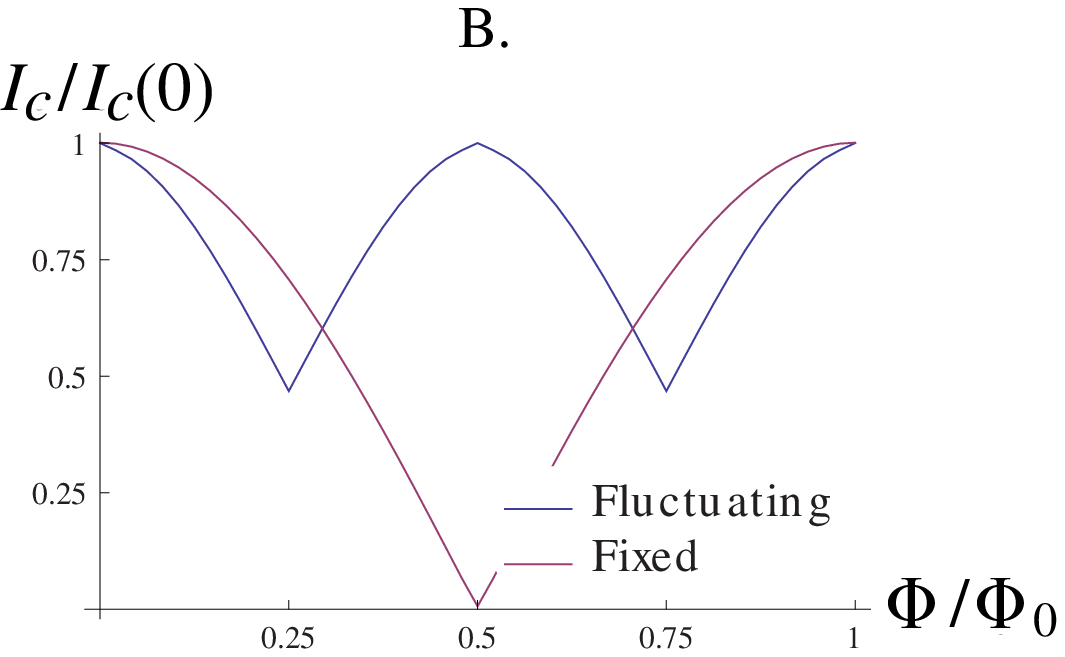}
\medskip
\includegraphics[width=1.7in]{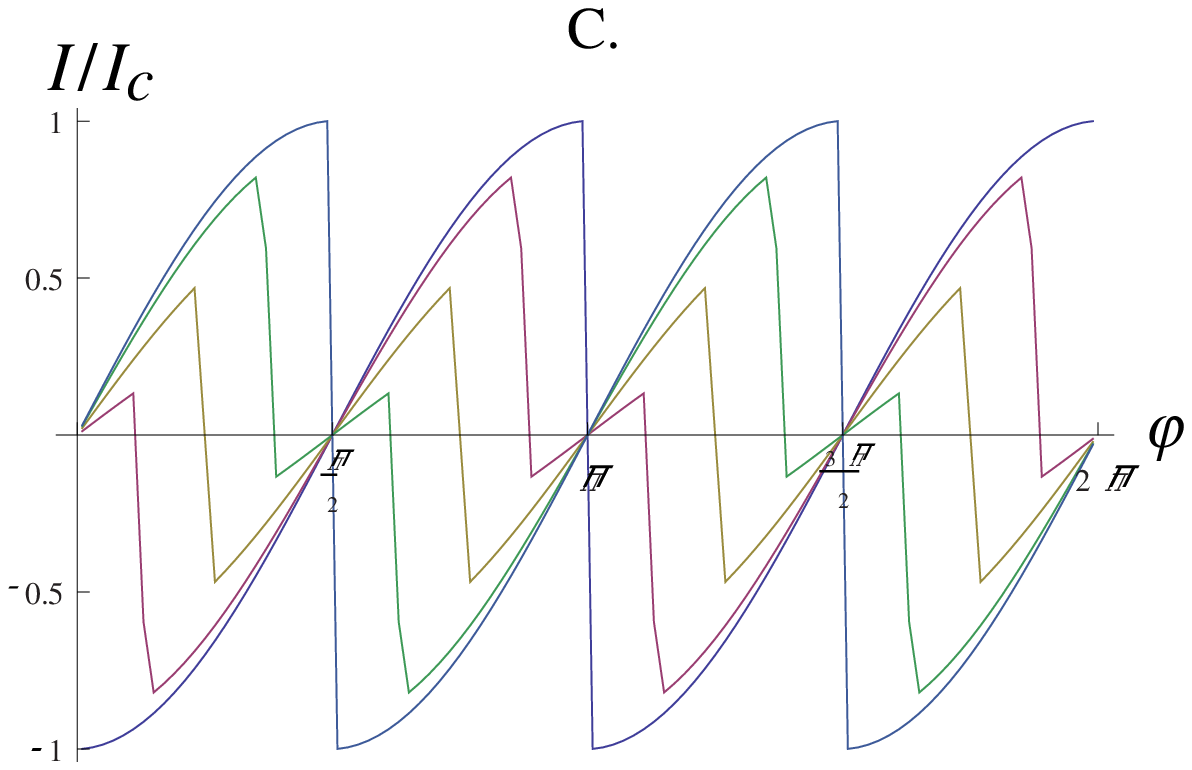}\includegraphics[width=1.7in]{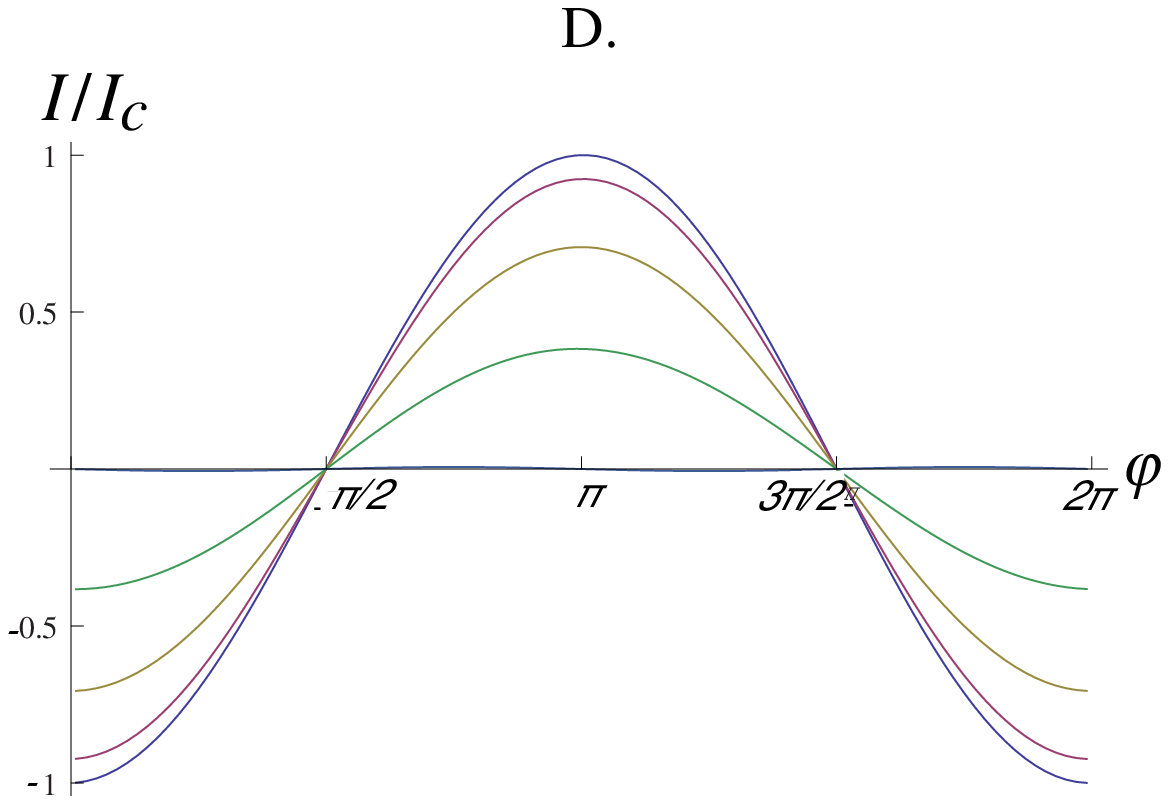}
\caption{A.: Schematic picture of a DC voltage-biased SQUID. %({\bf mark TSC and $s$-wave on the SQUID.}) 
B.: Dependence of the current oscillation amplitude on the flux $\Phi$. C. \& D.: Numerical calculation of the DC SQUID $I/I_c$ as a function of $\phi$ for the flux of $\Phi/\Phi_0 = 0,\frac{1}{8},\frac{1}{4},\frac{3}{8},\frac{1}{2}$ in blue, purple, brown, green, and black respectively; C. is for the slow limit and D. for the fast limit.}
\label{FIG:1d}
\end{figure}

To relate this behavior of Josephson junction to physical observable effects, we now consider a DC %superconducting quantum interference device 
SQUID built from two of such Josephson junctions. We now consider the TRI TSC chain to be finite and is coupled to the $s$-wave chain through hopping at the both ends of the chain, with flux $\Phi$ through the SQUID, as shown in Fig. 1 A. Each of two junctions has a Kramers doublet of Majorana zero modes $\gamma_{0\sigma}$ and $\gamma'_{0\sigma}$, where $\gamma'_{0\sigma} = -i(c_{N\sigma}-c^\dagger_{N\sigma})$ for the case of $\mu = 0, \Delta = t$. Therefore, the effective Hamiltonian of this SQUID is
\begin{equation}
H_{eff} = -J [N_{F1}\sin (\phi\!-\!\pi \Phi/\Phi_0) + N_{F2}\sin (\phi\!+\!\pi \Phi/\Phi_0)],
\end{equation}
where $\Phi_0 = h/2e$ is the unit quantum flux and $N_{F1}=i\gamma_{0\uparrow}\gamma_{0\downarrow}$, $N_{F2}=-i\gamma'_{0\uparrow}\gamma'_{0\downarrow}$ are two local fermion number parities. The behaviors of this SQUID %can be discussed
in the fast and slow limits are qualitatively different. In the slow limit $\omega_J\tau\ll 1$, the system stays at the lowest energy state with the energy $E_J = -J [|\sin (\phi - \pi \Phi/\Phi_0)|+|\sin (\phi + \pi \Phi/\Phi_0)|]$, so that the current-flux relation is
\begin{equation}
\frac{I}{I_c}= -\sum_{s=\pm} \frac{\sin \left(\phi + s\pi \frac{\Phi}{\Phi_0}\right)}{|\sin \left(\phi + s\pi \frac{\Phi}{\Phi_0}\right)|}\cos \left(\phi + s\pi \frac{\Phi}{\Phi_0}\right).
\end{equation}
As shown on the left plot of Fig. 1 B., there are sharp jumps in the current four times every period due to the fluctuation of the electron parity of each junction except at $\Phi=n\Phi_0$ and $\Phi=(n+1/2)\Phi_0$ where the electron parities of two junctions fluctuate together to give us the maximum current oscillation amplitude.

Due to the TR anomaly, %the behavior of 
whether this DC SQUID behaves like a normal SQUID or a $\pi$-SQUID in the fast limit $\omega_J \tau \gg 1$ is %strongly dependent on at what flux $\Phi$ we have ramped up the DC voltage. %, the DC SQUID can give us %a spontaneous flux as a consequence of the time-reversal anomaly. 
determined hysteretically. As we have Josephson oscillation without the fermion number parity fluctuation in the fast limit, we have four possible current-phase relations depending on the eigenvalues of $N_{F1}$, $N_{F2}$,
\begin{equation}
I =\left\{\begin{array}{c}-2 N_{F1} I_c\cos (\pi \Phi/\Phi_0) \cos \phi,~N_{F1}N_{F2}=+1\\
                                    -2 N_{F1} I_c\sin (\pi \Phi/\Phi_0) \sin \phi,~N_{F1}N_{F2}=-1\end{array}\right.
\end{equation}
We see that we have a normal SQUID for the even total fermion number parity $N_{F1}N_{F2}=+1$ and a $\pi$-SQUID with zero supercurrent at integer flux but maximal supercurrent at half-integer flux for the odd total fermion number parity $N_{F1}N_{F2}=-1$. %which implies %Unlike the $\pi$-SQUID consisting of the normal and $\pi$ junction, our $\pi$-SQUID have 
%at the zero flux a nonzero circulating current $I_s = \pm I_c/2$, and therefore %which would give us
%a spontaneous flux. %,   when the SQUID inductance is nonzero due to the TR anomaly of each junction. 
%By starting the Josephson oscillation at a low frequency and then speed up the frequency, it is possible obtain either the $\pi$ or the normal SQUID. %depending on whether the flux is integer or half-integer, for at these flux values, 
We now note that even in the slow limit, the SQUID is in the $N_{F1}N_{F2} = +1$ eigenstate at $\Phi=n\Phi_0$ and the $N_{F1}N_{F2} = +-1$ eigenstate at $\Phi=(n+1/2)\Phi_0$ as $N_{F1}$ and $N_{F2}$ fluctuate together at these flux values. %For instance, the $\pi$-SQUID can be obtained by starting the Josephson oscillation in the slow limit and then speeding up to the fast limit. Thus, %while we will see the current oscillating at the maximum amplitude when we ramp up the voltage while maintaining the flux at $\Phi = \Phi_0/2$, if we then turn off the flux to $\Phi = 0$ while maintaining the high voltage, the current oscillation will be completely turned off. %{\bf (Why first maximal oscillation?)}
Therefore we can obtain the $\pi$-SQUID is by ramping up the DC voltage from the slow to the fast limit while the flux is fixed at $\Phi = \Phi_0/2$, the normal SQUID by going through the same process at zero flux.
%while ramping up the voltage at the zero flux will give us the normal SQUID.

\begin{figure}
\includegraphics[width=1.6in]{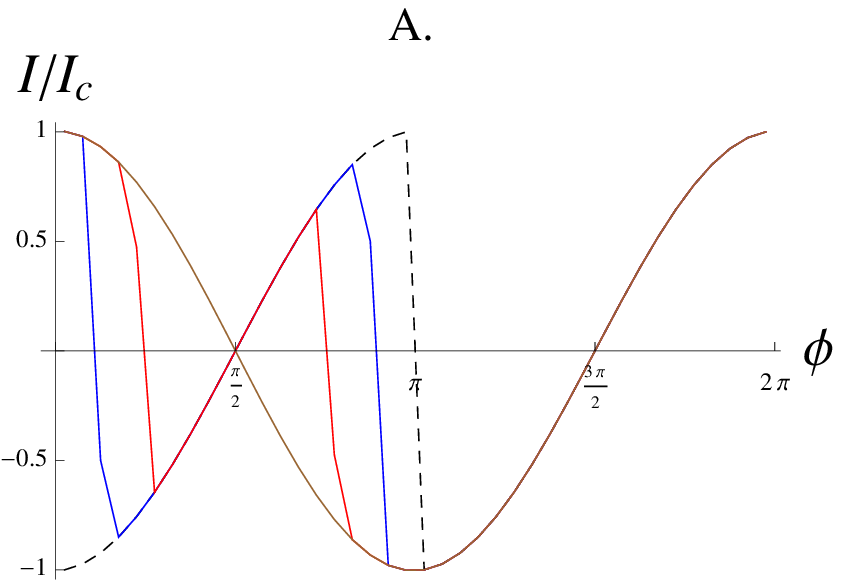}\includegraphics[width=1.6in]{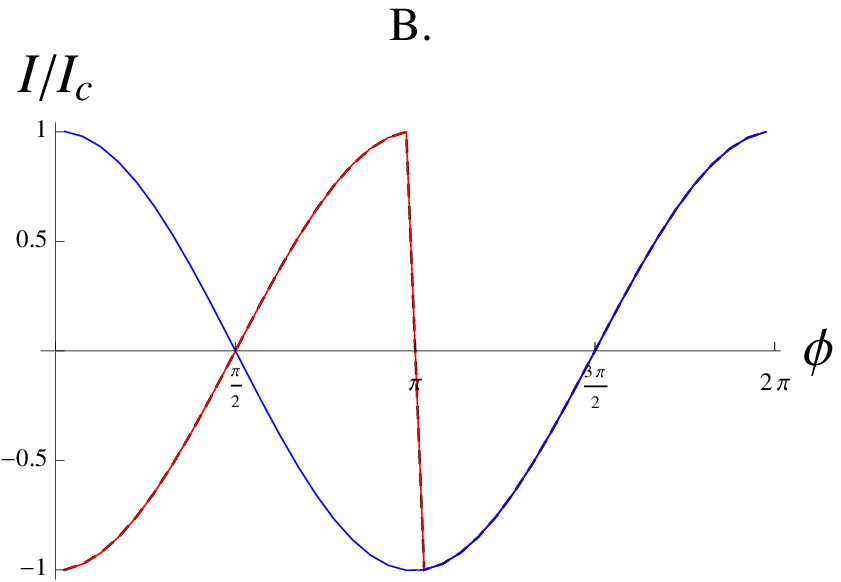}
\caption{The Zeeman field effect on the Josephson current in the slow limit for $\delta t/\Delta = 0.2$. %and the $s$-wave bandwidth of $w_s = 2|\Delta'| = 2\Delta$.
A.: The Zeeman field is applied along the junction with the magnitude of $g\mu_B h/\Delta$ = 0 (black dashed), 0.005 (blue), 0.01 (red), and 0.02 (brown). B.: The Zeeman field of magnitude $g\mu_B h/\Delta=0.02$ applied along the junction (red) and perpendicular to the junction (blue). While the field along the junction changes the location of discontinuity and eventually removes it altogether, the field perpendicular to junction has negligible effects. %({\bf explain the two curves in the right panel})
}
\label{FIG:1dZeeman}
\end{figure}

The effect of Zeeman field also reveals the unconventional nature of this Josephson junction. The Zeeman field breaks time reversal invariance and provides an alternative channel for gapping out the Majorana zero modes. To the leading order, our effective Hamiltonian of a single junction is modified to 
\begin{equation}
H_{eff} = (i\gamma_{0\uparrow}\gamma_{0\downarrow}) (J\sin \phi + g\mu_B {\bf h}\cdot{\bf \hat{n}}),
\label{EQ:1DjjZeeman}
\end{equation}
where $g$ is the $g$-factor of the normal state of TRI TSC, $\mu_B$ the Bohr magneton, and ${\bf h}$ is the effective Zeeman field, and ${\bf \hat{n}}$ is a unit vector along the junction \cite{chung2009, shindou2010}. %; we can see the sensitivity to the direction of the Zeeman field in the numerical results shown in Fig. \ref{FIG:1dZeeman}.  %In the simple model Hamiltonian (\ref{EQ:1Dtsc}) ${\bf \hat{n}}$ is the direction along the junction.
We can see that there is no level crossing between the even and odd parity states once ${\bf h}\cdot{\bf \hat{n}} > J/(g\mu_B)$, giving us a normal Josephson junction with no Majorana zero modes. The numerical calculation shown in Fig. \ref{FIG:1dZeeman} confirms both the sensitivity to the Zeeman field direction and the absence of level crossing at sufficiently strong field. Lastly, we note that the effect the Zeeman field adds to the Josephson coupling implies both the spin accumulation at the junction\cite{Kwon2004, Lu2009} and the Zeeman field induced $s$-wave pairing on the boundary of the TRI TSC. We will show that the same effects exist when the Majorana boundary states gets gapped in the higher dimensions. %({\bf I don't understand the last sentence. })%XLnote

\noindent{\bf Generalization to higher dimensions}

In higher dimensions, we can still detect through the Josephson effect a Zeeman field induced $s$-wave pairing at the boundary of a TRI TSC. This effect does not come from the Majorana zero modes; %fermion number parity, which in higher dimensions can be easily violated as the Majorana edge states of a TRI TSC has macroscopic number of modes forming a continuous spectrum, effectively always putting us
in fact, in the higher dimensions we are always in the slow limit $\omega_J\tau \ll 1$ as the boundary state of of a TRI TSC has macroscopic number of modes forming a continuous spectrum. However, the Josephson effect can be studied by considering a weak tunneling between TSC and an $s$-wave superconductor just as in 1D. For 2D the tunneling Hamiltonian can be written as
%\begin{equation}
$H_1 = - \sum_{n=1}^{N_y}\sum_\sigma[(\delta t)c^\dagger_\sigma (1,n)f_\sigma (m,n) + {\rm h.c.}]$
%\label{EQ:interHop2}
%\end{equation}
($n=1,2,...,N_y$ labels the boundary sites), and it %The tunneling Hamiltonian 
has a similar form for 3D. In both 2D and 3D, this coupling can only open a gap when the relative phase $\phi\neq 0,\pi$ just as in 1D.

%\subsection{Perturbation theory}

%{\bf Fig. 2. (a) $I-\phi$ relation of a single junction. (b) Fraunhofer pattern with and without in-plane field. (c) $I-\Phi$ relation of SQUID. (d) SOC effect to the $I-\Phi$ relation.}

%(In high D we assume we are always in the slow limit.)

\begin{figure}
\includegraphics[width=1.7in]{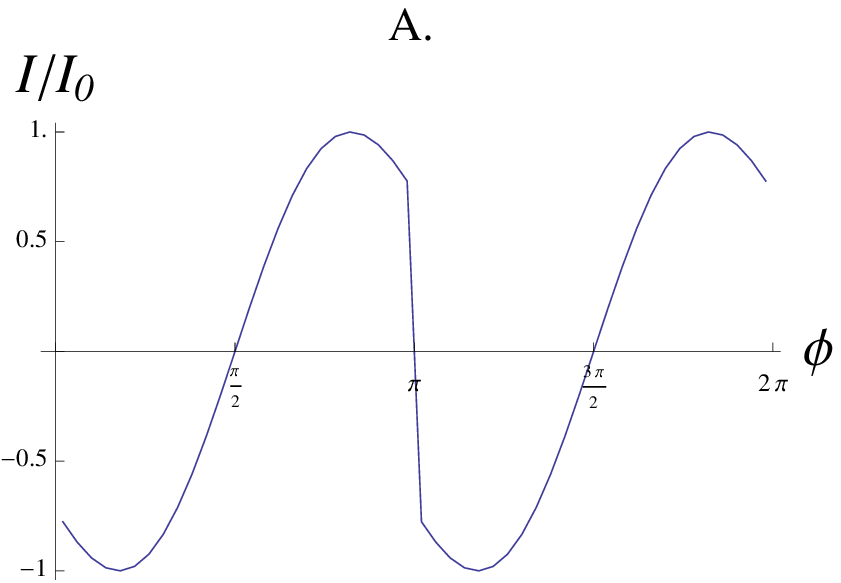}%\includegraphics[width=1.5in]{2x.eps}
\includegraphics[width=1.7in]{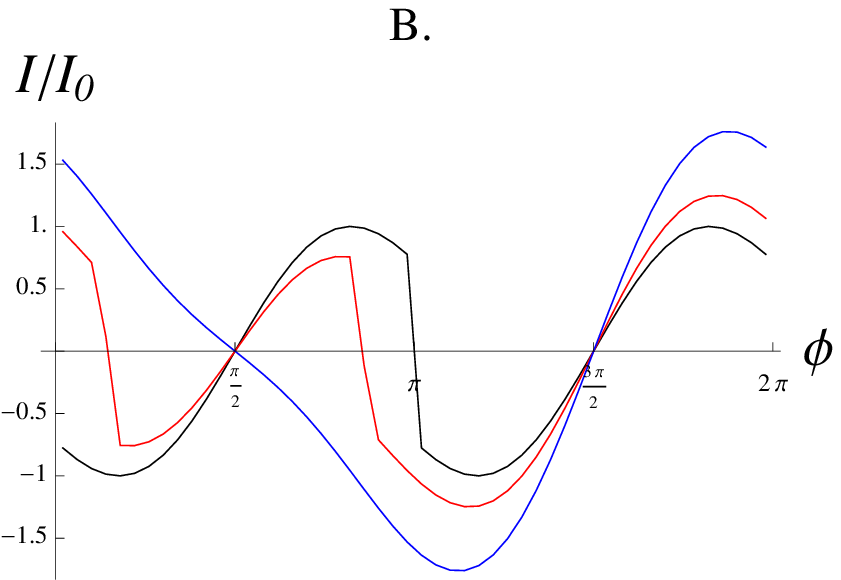}
\medskip
\includegraphics[width=1.65in]{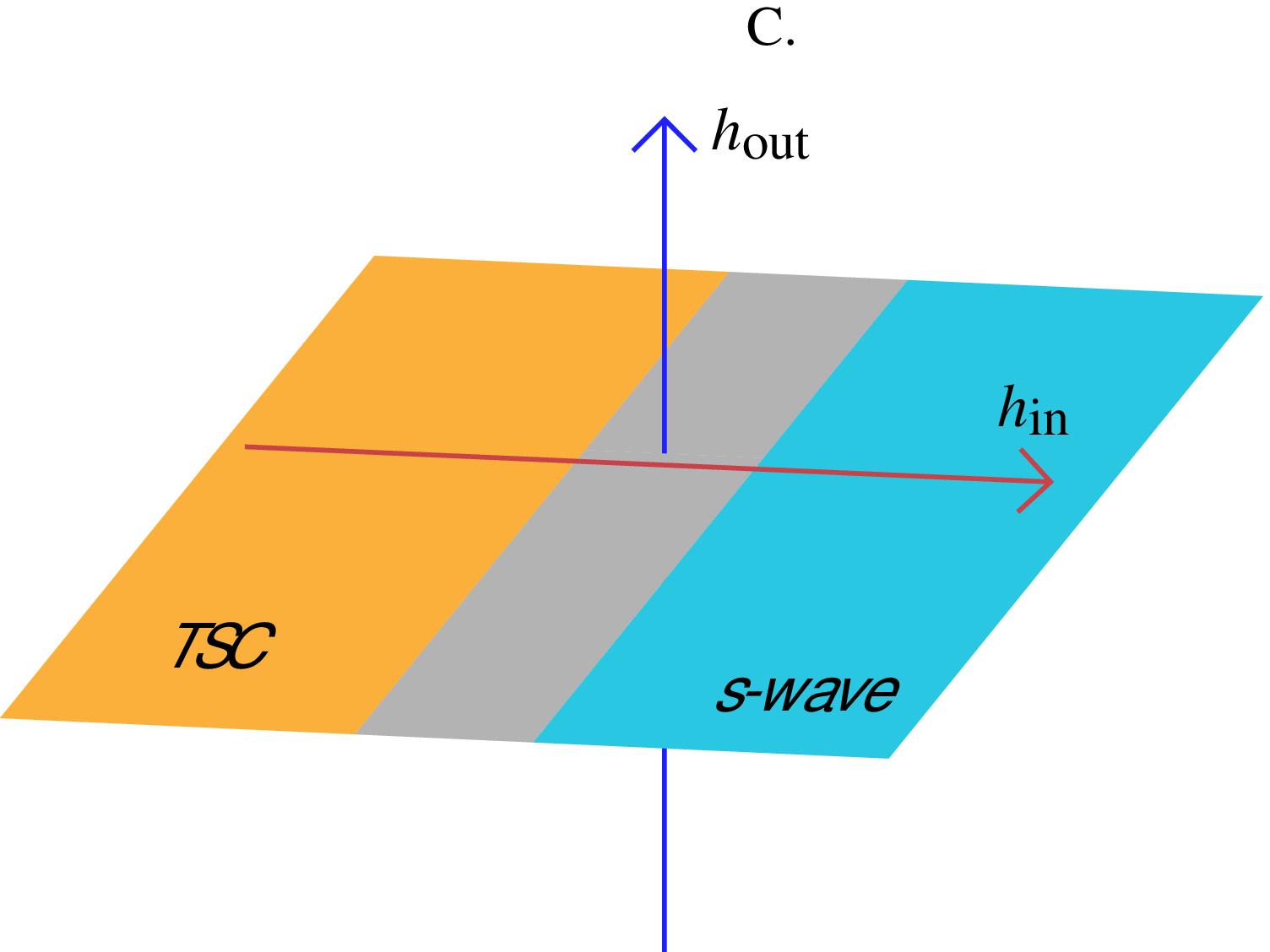}\includegraphics[width=1.7in]{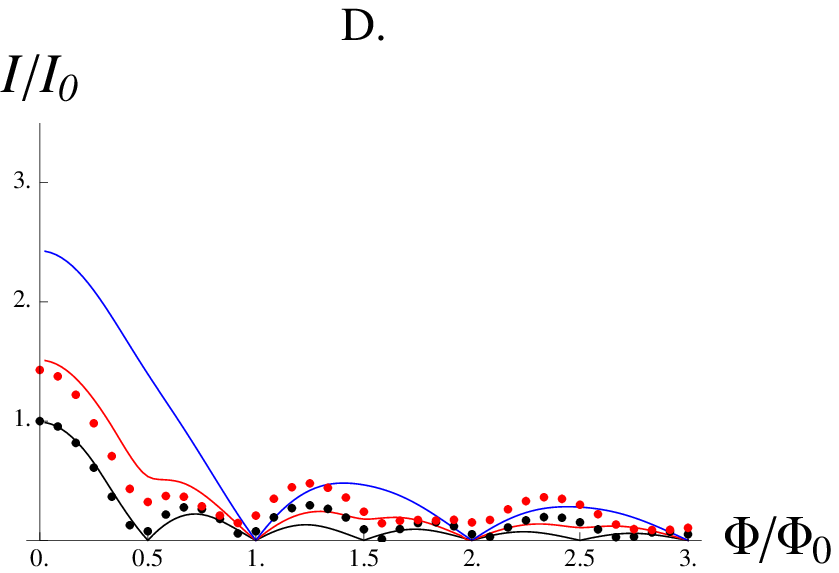}
\caption{We have set $\delta t/\Delta = 0.7$. %and $w_s = 6|\Delta'| = 6\Delta$
A.: Numerical results for the current-phase relation $I(\phi)$. %(b) the edge state dispersion near $k=0$ for $\phi=0$ (blue) and $\phi=\pi/2$ (red). 
B.: %The numerically calculation of 
Numerical results for $I(\phi)$ with the Zeeman field $g\mu_B h/\Delta =$ 0 (black), 0.02 (red), 0.06 (blue) along the junction. C.: The scheme for the Fraunhofer diffraction experiment. $h_{\rm out}$ gives us the flux $\Phi$ through the junction, $h_{\rm in}$ gives rise to Eq.(\ref{EQ:Zeeman2D}). D.: The Fraunhofer diffraction pattern with the Zeeman field $g\mu_B h/\Delta =$ 0 (black), 0.02 (red), 0.06 (blue) along the junction; the filled curves calculated from the current-phase relation Eq. (\ref{EQ:ZeemanJJ2d}) and the dots to the numerical calculation. %({\bf Josh, can you plot dispersion in a smaller range of $k$, such as 20 or 30 points in the range of $-\pi/50,\pi/50$. That should look better. })
}
\label{FIG:2d}
\end{figure}

The period doubling of the Josephson coupling in the higher dimension can be obtained for the weak tunneling limit by the second order perturbation theory exactly analogous to that of the 1D case. Here we will omit the detail of the perturbation expansion and directly write down the effective edge theory induced by the coupling. For the 1D Majorana edge state of 2D TSC, the effective Hamiltonian is
\begin{equation}
H_{\rm eff}=\sum_k\!\left[vk\left(\gamma_{-k\uparrow }\gamma_{k\uparrow}\!-\!\gamma_{-k\downarrow }\gamma_{k\downarrow}\right)\!+\!i(M\sin\phi)\gamma_{-k\downarrow}\gamma_{k\uparrow}\right].\nonumber\\
\label{EQ:Heff2D}
\end{equation}
As required by TRS, the mass term is proportional to $\sin\phi$; in fact it is roughly the same as the 1D Josephson coupling and therefore also proportional to $(\delta t)^2$. The 2D Josephson energy is determined by the ground state energy of the effective model (\ref{EQ:Heff2D}) defined with respect to that of the $M=0$ system:
\begin{eqnarray}
E_J(\phi)=\frac{1}{2}\sum_k\left[-\sqrt{v^2k^2+M^2\sin^2\phi}+v|k|\right]
\label{EQ:EJ2D}
\end{eqnarray}
In the thermodynamic limit the sum becomes an integral and we obtain the energy and current phase relation
\begin{eqnarray}
E_J(\phi)&=&\frac{\Phi_0 I_c}{2\pi}\left[1+\log\frac{4|\Delta|^2}{M^2\sin^2\phi}\right]\sin^2\phi,\nonumber\\
I(\phi) &=& -I_c \left[\log\frac{4|\Delta|^2}{M^2\sin^2\phi}\right]\sin 2\phi,
\label{EQ:2Djc}
\end{eqnarray}
where $I_c\propto M^2$. While the current-phase relation does not have discontinuity as in 1D, it does have a logarithmic singularity at $\phi=0,\pi$; this also shows up in the numerical calculation plotted in Fig. \ref{FIG:2d} A. As a consequence of the contribution from macroscopic number of modes, the Josephson current has a doubled frequency. Eq. (\ref{EQ:EJ2D}) remains valid for junctions between 3D TSC and $s$-wave superconductors, but there it gives us a 2D integral. Consequently, in the 3D case the junction has a doubled frequency without singularity: $I(\phi)=I_c\sin 2\phi$.

The in-plane Zeeman field in 2D can erase out this period doubling, thus affecting %a signature of
the TSC Josephson junction more dramatically than in 1D. Analogous to Eq.\ref{EQ:1DjjZeeman}, the Zeeman effect leads to an additional contribution to the gap of the edge state in the effective Hamiltonian (\ref{EQ:Heff2D})
\begin{equation}
H_Z =\sum_ki\gamma_{-k\downarrow}\gamma_{k\uparrow}g\mu_B {\bf h}\cdot{\bf \hat{n}}.
\label{EQ:Zeeman2D}
\end{equation}
with $\hat{\bf n}$ the unit vector along the edge normal. This modification will induce an ordinary $2\pi$ periodic Josephson coupling, since the current-phase relation is modified to
\begin{equation}
I(\phi) = -I_c \left[\log \frac{4|\Delta|^2/M^2}{(\sin \phi+ \tilde{h}/M)^2}\right]\left(\sin 2\phi+\frac{2\tilde{h}}{M}\cos\phi\right),
\label{EQ:ZeemanJJ2d}
\end{equation}
where $\tilde{h} \equiv g\mu_B {\bf h}\cdot{\bf \hat{n}}$. In fact this formula tells us that for $|\tilde{h}| \gg |M|$ our Josephson coupling becomes essentially conventional, with %the first harmonics dominating over all higher harmonics. %, up to the $\pi/2$ phase shift. %In addition, the singularity in the current will be removed for $|h_\parallel| > |\bar{M}|/g\mu_B$.
the effective critical current nearly proportional to the in-plane field. Given $|M|\propto (\delta t)^2$, this crossover occurs in the $|\tilde{h}| \ll \delta t$ regime, which is confirmed by the numerical calculation shown in Fig. \ref{FIG:2d} B. The underlying reason is that the Zeeman energy term of Eq.(\ref{EQ:Zeeman2D}) induces $s$-wave pairing correlation, which can be seen clearly in the real space basis. In fact, for the TRI TSC of Eq.(\ref{EQ:BWd}) we cannot gap out the Majorana boundary state without inducing the $s$-wave pairing correlation.% for $D=2,3$ as well as $D=1$.

This in-plane Zeeman field induced change in periodicity can be also seen in the Fraunhofer diffraction in a single Josephson junction. When the magnetic field is completely perpendicular, %Due to the doubled frequency of the Josephson current,
the Fraunhofer diffraction pattern for 2D %induced by a perpendicular magnetic field in the 2D Josephson junction
has zeros at both $n\Phi_0$ and $(n+1/2)\Phi_0$. %Similarly, a DC SQUID also has a doubled frequency in its current-flux relation.
We can see this by noting that the 2D Josephson current has only even harmonics and therefore can be expanded as $I(\phi, \Phi) = \sum_m I_{2m}[\sin (2\pi m\Phi/\Phi_0)/2\pi m\Phi/\Phi_0] \sin 2n\phi$. %This series expansion implies the suppression of the Josephson current at both integer and half-integer flux, for applying the flux $\Phi$ would lead to the modification $I_{2n} \to I_{2n} (\sin (2\pi n\Phi/\Phi_0)/2\pi n\Phi/\Phi_0)$. %in our , we have % through the junction.
However if one measures the Fraunhofer pattern in a canted magnetic field with both perpendicular and in-plane components, Eq.(\ref{EQ:ZeemanJJ2d}) gives us nonzero odd harmonics in the current phase relation as shown in Fig. \ref{FIG:2d} B., which leads to non-zero critical currents at half-integer fluxes as shown in Fig \ref{FIG:2d} (d). %proportional to $\tilde{h}$.
%The fact that in-plane magnetic field restores 
This restoration of the ordinary Josephson coupling %is dramatically different from ordinary Josephson junctions and thus 
provides a sharp experimental signature of the TSC.

%\section{Summary}

\noindent{\bf Summary}

We studied the topological Josephson coupling between the TRI TSC and the conventional superconductor for $D=1,2,3$ which is qualitatively different from ordinary Josephson effect. The effects proposed originate from the time-reversal anomaly--the anticommutation between the time reversal operation and the local fermion number parity of the topological edge/surface states.
%The main feature in 1D is %the time-reversal anomaly due to the anticommutation between the time reversal operation and the local fermion number parity, which leads to the nonzero Josephson current at $\phi=0,\pi$.
In 1D, the DC SQUID made of two such junctions can hysteretically behave either a normal or a $\pi$ SQUID. In 2D and 3D the Josephson coupling has a $\pi$ periodicity, leading to a half period Fraunhofer pattern and a half period DC SQUID. Since the topological coupling is determined by the time-reversal protected edge/surface states, a Zeeman magnetic field breaking time-reversal symmetry is predicted to restore the normal behavior in the junction, and induce a cross-over of the topological Josephson junction back to a normal one.

%A Zeeman field along the junction can induce a normal behavior in the junction, another evidence that time reversal symmetry plays a crucial role in this Josephson effect. We find that this Zeeman field effect persists on in 2D, where the Josephson effect has $\pi$ periodicity in when the Zeeman field along the junction is zero but the first harmonics become dominant when the Zeeman energy becomes larger than the $s$-wave induced gap. This Zeeman field induced change in the Josephson coupling periodicity is present for the 3D TRI TSC also, and therefore should provide a very sharp method to detect TSC.

\noindent{\bf Acknowledgement}

It is our pleasure to thank Mac Beasley, Srinivas Raghu and Shou-Cheng Zhang for sharing their insights with us. SBC, JH and XLQ are supported by the Department of Energy, Office
of Basic Energy Sciences, Division of Materials Sciences
and Engineering, under contract DE-AC02-76SF00515.

%JH by the Department of Physics, Stanford University, and XLQ by the Packard foundation and the Alfred P. Sloan foundation.

\bibliography{TI}

%Merlin.mbs v4.21 2009-07-09.
\begin{thebibliography}{10}%
\makeatletter
\providecommand \@ifxundefined [1]{%
 \ifx #1\undefined \expandafter \@firstoftwo
 \else \expandafter \@secondoftwo
\fi
}%
\providecommand \@ifnum [1]{%
 \ifnum #1\expandafter \@firstoftwo
 \else \expandafter \@secondoftwo
\fi
}%
\providecommand \enquote [1]{``#1''}%
\providecommand \bibnamefont  [1]{#1}%
\providecommand \bibfnamefont [1]{#1}%
\providecommand \citenamefont [1]{#1}%
\providecommand\href[0]{\@sanitize\@href}%
\providecommand\@href[1]{\endgroup\@@startlink{#1}\endgroup\@@href}%
\providecommand\@@href[1]{#1\@@endlink}%
\providecommand \@sanitize [0]{\begingroup\catcode`\&12\catcode`\#12\relax}%
\@ifxundefined \pdfoutput {\@firstoftwo}{%
 \@ifnum{\z@=\pdfoutput}{\@firstoftwo}{\@secondoftwo}%
}{%
 \providecommand\@@startlink[1]{\leavevmode\special{html:<a href="#1">}}%
 \providecommand\@@endlink[0]{\special{html:</a>}}%
}{%
 \providecommand\@@startlink[1]{%
  \leavevmode
  \pdfstartlink
   attr{/Border[0 0 1 ]/H/I/C[0 1 1]}%
   user{/Subtype/Link/A<</Type/Action/S/URI/URI(#1)>>}%
  \relax
 }%
 \providecommand\@@endlink[0]{\pdfendlink}%
}%
\providecommand \url  [0]{\begingroup\@sanitize \@url }%
\providecommand \@url [1]{\endgroup\@href {#1}{\urlprefix}}%
\providecommand \urlprefix [0]{URL }%
\providecommand \Eprint[0]{\href }%
\@ifxundefined \urlstyle {%
  \providecommand \doi [1]{doi:\discretionary{}{}{}#1}%
}{%
  \providecommand \doi [0]{doi:\discretionary{}{}{}\begingroup
  \urlstyle{rm}\Url }%
}%
\providecommand \doibase [0]{http://dx.doi.org/}%
\providecommand \Doi[1]{\href{\doibase#1}}%
\providecommand \bibAnnote [3]{%
  \BibitemShut{#1}%
  \begin{quotation}\noindent
    \textsc{Key:}\ #2\\\textsc{Annotation:}\ #3%
  \end{quotation}%
}%
\providecommand \bibAnnoteFile [2]{%
  \IfFileExists{#2}{\bibAnnote {#1} {#2} {\input{#2}}}{}%
}%
\providecommand \typeout [0]{\immediate \write \m@ne }%
\providecommand \selectlanguage [0]{\@gobble}%
\providecommand \bibinfo [0]{\@secondoftwo}%
\providecommand \bibfield [0]{\@secondoftwo}%
\providecommand \translation [1]{[#1]}%
\providecommand \BibitemOpen[0]{}%
\providecommand \bibitemStop [0]{}%
\providecommand \bibitemNoStop [0]{.\EOS\space}%
\providecommand \EOS [0]{\spacefactor3000\relax}%
\providecommand \BibitemShut [1]{\csname bibitem#1\endcsname}%
%</preamble>
\bibitem{qi2011rmp}%
  \BibitemOpen
  \bibfield{author}{%
  \bibinfo {author} {\bibfnamefont{X.-L.}\ \bibnamefont{Qi}}\ and\ \bibinfo
  {author} {\bibfnamefont{S.-C.}\ \bibnamefont{Zhang}},\ }%
  \bibfield{journal}{%
  \bibinfo {journal} {Rev. Mod. Phys.}\ }%
  \textbf{\bibinfo {volume} {83}},\ \bibinfo {pages} {1057} (\bibinfo {year}
  {2011})%
  \bibAnnoteFile{NoStop}{qi2011rmp}%
\bibitem{hasan2010}%
  \BibitemOpen
  \bibfield{author}{%
  \bibinfo {author} {\bibfnamefont{M.~Z.}\ \bibnamefont{Hasan}}\ and\ \bibinfo
  {author} {\bibfnamefont{C.~L.}\ \bibnamefont{Kane}},\ }%
  \bibfield{journal}{%
  \bibinfo {journal} {Rev. Mod. Phys.}\ }%
  \textbf{\bibinfo {volume} {82}},\ \bibinfo {pages} {3045} (\bibinfo {year}
  {2010})%
  \bibAnnoteFile{NoStop}{hasan2010}%
\bibitem{moore2010}%
  \BibitemOpen
  \bibfield{author}{%
  \bibinfo {author} {\bibfnamefont{J.~E.}\ \bibnamefont{Moore}},\ }%
  \bibfield{journal}{%
  \bibinfo {journal} {Nature}\ }%
  \textbf{\bibinfo {volume} {464}},\ \bibinfo {pages} {194} (\bibinfo {year}
  {2010})%
  \bibAnnoteFile{NoStop}{moore2010}%
\bibitem{read2000}%
  \BibitemOpen
  \bibfield{author}{%
  \bibinfo {author} {\bibfnamefont{N.}~\bibnamefont{Read}}\ and\ \bibinfo
  {author} {\bibfnamefont{D.}~\bibnamefont{Green}},\ }%
  \bibfield{journal}{%
  \bibinfo {journal} {Phys. Rev. B}\ }%
  \textbf{\bibinfo {volume} {61}},\ \bibinfo {pages} {10267} (\bibinfo {year}
  {2000})%
  \bibAnnoteFile{NoStop}{read2000}%
\bibitem{schnyder2008}%
  \BibitemOpen
  \bibfield{author}{%
  \bibinfo {author} {\bibfnamefont{A.~P.}\ \bibnamefont{Schnyder}}, \bibinfo
  {author} {\bibfnamefont{S.}~\bibnamefont{Ryu}}, \bibinfo {author}
  {\bibfnamefont{A.}~\bibnamefont{Furusaki}},\ and\ \bibinfo {author}
  {\bibfnamefont{A.~W.~W.}\ \bibnamefont{Ludwig}},\ }%
  \bibfield{journal}{%
  \bibinfo {journal} {Phys. Rev. B}\ }%
  \textbf{\bibinfo {volume} {78}},\ \bibinfo {pages} {195125} (\bibinfo {year}
  {2008})%
  \bibAnnoteFile{NoStop}{schnyder2008}%
\bibitem{roy2008}%
  \BibitemOpen
  \bibfield{author}{%
  \bibinfo {author} {\bibfnamefont{R.}~\bibnamefont{Roy}},\ }%
  \bibinfo {howpublished} {e-print arXiv:0803.2868} (\bibinfo {year} {2008})%
  \bibAnnoteFile{NoStop}{roy2008}%
\bibitem{qi2009b}%
  \BibitemOpen
  \bibfield{author}{%
  \bibinfo {author} {\bibfnamefont{X.-L.}\ \bibnamefont{Qi}}, \bibinfo {author}
  {\bibfnamefont{T.~L.}\ \bibnamefont{Hughes}}, \bibinfo {author}
  {\bibfnamefont{S.}~\bibnamefont{Raghu}},\ and\ \bibinfo {author}
  {\bibfnamefont{S.-C.}\ \bibnamefont{Zhang}},\ }%
  \bibfield{journal}{%
  \bibinfo {journal} {Phys. Rev. Lett.}\ }%
  \textbf{\bibinfo {volume} {102}},\ \bibinfo {pages} {187001} (\bibinfo {year}
  {2009})%
  \bibAnnoteFile{NoStop}{qi2009b}%
\bibitem{chung2009}%
  \BibitemOpen
  \bibfield{author}{%
  \bibinfo {author} {\bibfnamefont{S.~B.}\ \bibnamefont{Chung}}\ and\ \bibinfo
  {author} {\bibfnamefont{S.~C.}\ \bibnamefont{Zhang}},\ }%
  \bibfield{journal}{%
  \bibinfo {journal} {Phys. Rev. Lett.}\ }%
  \textbf{\bibinfo {volume} {103}},\ \bibinfo {pages} {235301} (\bibinfo {year}
  {2009})%
  \bibAnnoteFile{NoStop}{chung2009}%
\bibitem{fu2010b}%
  \BibitemOpen
  \bibfield{author}{%
  \bibinfo {author} {\bibfnamefont{L.}~\bibnamefont{Fu}}\ and\ \bibinfo
  {author} {\bibfnamefont{E.}~\bibnamefont{Berg}},\ }%
  \bibfield{journal}{%
  \Doi{10.1103/PhysRevLett.105.097001}{\bibinfo {journal} {Phys. Rev. Lett.}}\
  }%
  \textbf{\bibinfo {volume} {105}},\ \bibinfo {pages} {097001} (\bibinfo {year}
  {2010})%
  \bibAnnoteFile{NoStop}{fu2010b}%
\bibitem{hsieh2012}%
  \BibitemOpen
  \bibfield{author}{%
  \bibinfo {author} {\bibfnamefont{T.~H.}\ \bibnamefont{Hsieh}}\ and\ \bibinfo
  {author} {\bibfnamefont{L.}~\bibnamefont{Fu}},\ }%
  \bibfield{journal}{%
  \Doi{10.1103/PhysRevLett.108.107005}{\bibinfo {journal} {Phys. Rev. Lett.}}\
  }%
  \textbf{\bibinfo {volume} {108}},\ \bibinfo {pages} {107005} (\bibinfo {year}
  {2012})%
  \bibAnnoteFile{NoStop}{hsieh2012}%
\bibitem{sasaki2011}%
  \BibitemOpen
  \bibfield{author}{%
  \bibinfo {author} {\bibfnamefont{S.}~\bibnamefont{Sasaki}}, \bibinfo {author}
  {\bibfnamefont{M.}~\bibnamefont{Kriener}}, \bibinfo {author}
  {\bibfnamefont{K.}~\bibnamefont{Segawa}}, \bibinfo {author}
  {\bibfnamefont{K.}~\bibnamefont{Yada}}, \bibinfo {author}
  {\bibfnamefont{Y.}~\bibnamefont{Tanaka}}, \bibinfo {author}
  {\bibfnamefont{M.}~\bibnamefont{Sato}},\ and\ \bibinfo {author}
  {\bibfnamefont{Y.}~\bibnamefont{Ando}},\ }%
  \bibfield{journal}{%
  \Doi{10.1103/PhysRevLett.107.217001}{\bibinfo {journal} {Phys. Rev. Lett.}}\
  }%
  \textbf{\bibinfo {volume} {107}},\ \bibinfo {pages} {217001} (\bibinfo {year}
  {2011})%
  \bibAnnoteFile{NoStop}{sasaki2011}%
\bibitem{yan2010a}%
  \BibitemOpen
  \bibfield{author}{%
  \bibinfo {author} {\bibfnamefont{B.}~\bibnamefont{Yan}}, \bibinfo {author}
  {\bibfnamefont{C.-X.}\ \bibnamefont{Liu}}, \bibinfo {author}
  {\bibfnamefont{H.}~\bibnamefont{Zhang}}, \bibinfo {author}
  {\bibfnamefont{C.~Y.}\ \bibnamefont{Yam}}, \bibinfo {author}
  {\bibfnamefont{X.~L.}\ \bibnamefont{Qi}}, \bibinfo {author}
  {\bibfnamefont{T.}~\bibnamefont{Frauenheim}},\ and\ \bibinfo {author}
  {\bibfnamefont{S.~C.}\ \bibnamefont{Zhang}},\ }%
  \bibfield{journal}{%
  \bibinfo {journal} {Europhys. Lett.}\ }%
  \textbf{\bibinfo {volume} {90}},\ \bibinfo {pages} {37002} (\bibinfo {year}
  {2010})%
  \bibAnnoteFile{NoStop}{yan2010a}%
\bibitem{niemi1983}%
  \BibitemOpen
  \bibfield{author}{%
  \bibinfo {author} {\bibfnamefont{A.~J.}\ \bibnamefont{Niemi}}\ and\ \bibinfo
  {author} {\bibfnamefont{G.~W.}\ \bibnamefont{Semenoff}},\ }%
  \bibfield{journal}{%
  \bibinfo {journal} {Phys. Rev. Lett.}\ }%
  \textbf{\bibinfo {volume} {51}},\ \bibinfo {pages} {2077} (\bibinfo {year}
  {1983})%
  \bibAnnoteFile{NoStop}{niemi1983}%
\bibitem{redlich1984}%
  \BibitemOpen
  \bibfield{author}{%
  \bibinfo {author} {\bibfnamefont{A.~N.}\ \bibnamefont{Redlich}},\ }%
  \bibfield{journal}{%
  \bibinfo {journal} {Phys. Rev. D}\ }%
  \textbf{\bibinfo {volume} {29}},\ \bibinfo {pages} {2366} (\bibinfo {year}
  {1984})%
  \bibAnnoteFile{NoStop}{redlich1984}%
\bibitem{wang2011}%
  \BibitemOpen
  \bibfield{author}{%
  \bibinfo {author} {\bibfnamefont{Z.}~\bibnamefont{Wang}}, \bibinfo {author}
  {\bibfnamefont{X.-L.}\ \bibnamefont{Qi}},\ and\ \bibinfo {author}
  {\bibfnamefont{S.-C.}\ \bibnamefont{Zhang}},\ }%
  \bibfield{journal}{%
  \bibinfo {journal} {Phys. Rev. B}\ }%
  \textbf{\bibinfo {volume} {84}},\ \bibinfo {pages} {014527} (\bibinfo {year}
  {2011})%
  \bibAnnoteFile{NoStop}{wang2011}%
\bibitem{ryu2012b}%
  \BibitemOpen
  \bibfield{author}{%
  \bibinfo {author} {\bibfnamefont{S.}~\bibnamefont{Ryu}}, \bibinfo {author}
  {\bibfnamefont{J.~E.}\ \bibnamefont{Moore}},\ and\ \bibinfo {author}
  {\bibfnamefont{A.~W.~W.}\ \bibnamefont{Ludwig}},\ }%
  \bibfield{journal}{%
  \Doi{10.1103/PhysRevB.85.045104}{\bibinfo {journal} {Phys. Rev. B}}\ }%
  \textbf{\bibinfo {volume} {85}},\ \bibinfo {pages} {045104} (\bibinfo {year}
  {2012})%
  \bibAnnoteFile{NoStop}{ryu2012b}%
\bibitem{qi2012b}%
  \BibitemOpen
  \bibfield{author}{%
  \bibinfo {author} {\bibfnamefont{X.-L.}\ \bibnamefont{Qi}}, \bibinfo {author}
  {\bibfnamefont{E.}~\bibnamefont{Witten}},\ and\ \bibinfo {author}
  {\bibfnamefont{S.-C.}\ \bibnamefont{Zhang}},\ }%
  \bibinfo {howpublished} {e-print arXiv:1206.1407}%
  \bibAnnoteFile{NoStop}{qi2012b}%
\bibitem{Note1}%
  \BibitemOpen
  \bibinfo {note} {It should be clarified that the topological Josephson effect
  discussed in this work is qualitatively different from the fractional
  Josephson effect discussed in Ref. \cite {fu2008} which applies to a
  different physical system and has a halved rather than doubled frequency.}%
  \bibAnnoteFile{Stop}{Note1}%
\bibitem{lee1997}%
  \BibitemOpen
  \bibfield{author}{%
  \bibinfo {author} {\bibfnamefont{I.~J.}\ \bibnamefont{Lee}}, \bibinfo
  {author} {\bibfnamefont{M.~J.}\ \bibnamefont{Naughton}}, \bibinfo {author}
  {\bibfnamefont{G.~M.}\ \bibnamefont{Danner}},\ and\ \bibinfo {author}
  {\bibfnamefont{P.~M.}\ \bibnamefont{Chaikin}},\ }%
  \bibfield{journal}{%
  \Doi{10.1103/PhysRevLett.78.3555}{\bibinfo {journal} {Phys. Rev. Lett.}}\ }%
  \textbf{\bibinfo {volume} {78}},\ \bibinfo {pages} {3555} (\bibinfo {year}
  {1997})%
  \bibAnnoteFile{NoStop}{lee1997}%
\bibitem{mercure2012}%
  \BibitemOpen
  \bibfield{author}{%
  \bibinfo {author} {\bibfnamefont{J.-F.}\ \bibnamefont{Mercure}}, \bibinfo
  {author} {\bibfnamefont{A.~F.}\ \bibnamefont{Bangura}}, \bibinfo {author}
  {\bibfnamefont{X.}~\bibnamefont{Xu}}, \bibinfo {author}
  {\bibfnamefont{N.}~\bibnamefont{Wakeham}}, \bibinfo {author}
  {\bibfnamefont{A.}~\bibnamefont{Carrington}}, \bibinfo {author}
  {\bibfnamefont{P.}~\bibnamefont{Walmsley}}, \bibinfo {author}
  {\bibfnamefont{M.}~\bibnamefont{Greenblatt}},\ and\ \bibinfo {author}
  {\bibfnamefont{N.~E.}\ \bibnamefont{Hussey}},\ }%
  \bibfield{journal}{%
  \Doi{10.1103/PhysRevLett.108.187003}{\bibinfo {journal} {Phys. Rev. Lett.}}\
  }%
  \textbf{\bibinfo {volume} {108}},\ \bibinfo {pages} {187003} (\bibinfo {year}
  {2012})%
  \bibAnnoteFile{NoStop}{mercure2012}%
\bibitem{kitaev2001}%
  \BibitemOpen
  \bibfield{author}{%
  \bibinfo {author} {\bibfnamefont{A.~Y.}\ \bibnamefont{Kitaev}},\ }%
  \bibfield{journal}{%
  \bibinfo {journal} {Physics-Uspekhi}\ }%
  \textbf{\bibinfo {volume} {44}},\ \bibinfo {pages} {131} (\bibinfo {year}
  {2001})%
  \bibAnnoteFile{NoStop}{kitaev2001}%
\bibitem{shindou2010}%
  \BibitemOpen
  \bibfield{author}{%
  \bibinfo {author} {\bibfnamefont{R.}~\bibnamefont{Shindou}}, \bibinfo
  {author} {\bibfnamefont{A.}~\bibnamefont{Furusaki}},\ and\ \bibinfo {author}
  {\bibfnamefont{N.}~\bibnamefont{Nagaosa}},\ }%
  \bibfield{journal}{%
  \Doi{10.1103/PhysRevB.82.180505}{\bibinfo {journal} {Phys. Rev. B}}\ }%
  \textbf{\bibinfo {volume} {82}},\ \bibinfo {pages} {180505} (\bibinfo {year}
  {2010})%
  \bibAnnoteFile{NoStop}{shindou2010}%
\bibitem{Kwon2004}%
  \BibitemOpen
  \bibinfo {howpublished} {H.-J.~Kwon, K.~Sengupta, and V.~M.~Yakovenko, Eur.\
  Phys.\ J. B {\bf 37}, 349 (2004); K.~Sengupta, and V.~M.~Yakovenko, Phys.\
  Rev.\ Lett. {\bf 101}, 187003 (2008)}%
  \bibAnnoteFile{NoStop}{Kwon2004}%
\bibitem{Lu2009}%
  \BibitemOpen
  \bibinfo {howpublished} {C.-K.~Lu and S.~Yip, Phys.\ Rev.\ B {\bf 80}, 024504
  (2009); Z.~Yang, J.~Wang, K.~S.~Chan, J.\ Phys.:\ Condens.\ Matter {\bf 23}
  085701 (2011)}%
  \bibAnnoteFile{NoStop}{Lu2009}%
\bibitem{fu2008}%
  \BibitemOpen
  \bibfield{author}{%
  \bibinfo {author} {\bibfnamefont{L.}~\bibnamefont{Fu}}\ and\ \bibinfo
  {author} {\bibfnamefont{C.~L.}\ \bibnamefont{Kane}},\ }%
  \bibfield{journal}{%
  \bibinfo {journal} {Phys. Rev. Lett.}\ }%
  \textbf{\bibinfo {volume} {100}},\ \bibinfo {pages} {096407} (\bibinfo {year}
  {2008})%
  \bibAnnoteFile{NoStop}{fu2008}%
\bibitem{asano2003}%
  \BibitemOpen
  \bibfield{author}{%
  \bibinfo {author} {\bibfnamefont{Y.}~\bibnamefont{Asano}}, \bibinfo {author}
  {\bibfnamefont{Y.}~\bibnamefont{Tanaka}}, \bibinfo {author}
  {\bibfnamefont{M.}~\bibnamefont{Sigrist}},\ and\ \bibinfo {author}
  {\bibfnamefont{S.}~\bibnamefont{Kashiwaya}},\ }%
  \bibfield{journal}{%
  \Doi{10.1103/PhysRevB.67.184505}{\bibinfo {journal} {Phys. Rev. B}}\ }%
  \textbf{\bibinfo {volume} {67}},\ \bibinfo {pages} {184505} (\bibinfo {year}
  {2003})%
  \bibAnnoteFile{NoStop}{asano2003}%
\end{thebibliography}%

\appendix

\section{Models of TSC in $D=1,2$.}
%For later convenience we also write down explicitly the Hamiltonian for $D=1$ and $D=2$. For $D=1$, the Hamiltonian describes two copies of spinless $p$-wave superconductors with the same pairing gap for spin up-up pairs and down-down pairs,
Here we present the explicit form of the Hamiltonian (\ref{EQ:BWd}) for $D=1$ and $D=2$. For our numerical calculation for $D=1$ in Figs. 1 and 2, the BdG Hamiltonians for our two superconducting chains are
\begin{eqnarray}
H_{p} =&-&\mu \sum_{{\bf r}s}c^\dagger_{{\bf r}s}c_{{\bf r}s} - t \sum_{{\bf r}s}(c^\dagger_{{\bf r}+{\bf \hat{x}} s}c_{{\bf r}s} + {\rm h.c.})\nonumber\\
&+& \sum_{{\bf r}s}(\Delta c^\dagger_{{\bf r}+{\bf \hat{x}} s}c^\dagger_{{\bf r}s} + {\rm h.c.}),
\label{EQ:1Dtsc}
\end{eqnarray}
for the TRI TSC and
\begin{eqnarray}
H_s =&-&\mu' \sum_{{\bf r'}s} f^\dagger_{{\bf r'}s} f_{{\bf r'}s} - t'\sum_{{\bf r'}s} (f^\dagger_{{\bf r'}+{\bf \hat{x}} s} f_{{\bf r'}s}+ {\rm h.c.})\nonumber\\
&+&  \sum_{\bf r'}(\Delta' f^\dagger_{{\bf r'}\uparrow} f^\dagger_{{\bf r'}\downarrow} + {\rm h.c.}).
\end{eqnarray}
for the $s$-wave chain; note that Eq.(\ref{EQ:1Dtsc}) can be obtained from the $D=1$ case of Eq.(\ref{EQ:BWd}) by applying $-\pi/2$ spin rotation around the $z$ axis. We have set $|t|=|t'|=|\Delta|=|\Delta'|$, $\mu_s = \mu_p = 0$, $|\delta t/t|=0.2$, and both chains consisted of 100 lattice sites. For the numerical calculation in Fig. 2, we add the Zeeman field term
\begin{equation}
H_Z = -g\mu_B \sum_{ss'} ({\bf h}\cdot{\bm\sigma})_{ss'} \left[\sum_{\bf r} c^\dagger_{{\bf r}s}c_{{\bf r}s'} + \sum_{\bf r'}  f^\dagger_{{\bf r'}s} f_{{\bf r'}s'}\right].
\end{equation}
%We point out that the $y$-direction in the spin basis of Eq.(\ref{EQ:1Dtsc}) is the direction along the junction for
%The TSC phase has a Kramer's doublet of Majorana zero modes, $\gamma_{0\uparrow}, \gamma_{0\downarrow}$ at the boundary. For $D=2$ the Hamiltonian has the form of

For the $D=2$ numerical calculation in Fig. \ref{FIG:2d}, we used for the TRI TSC the BdG Hamiltonian
\begin{eqnarray}
H_{p} =&-&\mu \sum_{{\bf r}s}c^\dagger_{{\bf r}s}c_{{\bf r}s} - t \sum_{{\bf \hat{e}}={\bf \hat{x}},{\bf \hat{y}}}\sum_{{\bf r}s}(c^\dagger_{{\bf r}+{\bf \hat{e}} s}c_{{\bf r}s} + {\rm h.c.})\nonumber\\
&+&  \Delta\sum_{{\bf r}s}[(c^\dagger_{{\bf r}+{\bf \hat{x}} s}c^\dagger_{{\bf r}s}-isc^\dagger_{{\bf r}+{\bf \hat{y}} s}c^\dagger_{{\bf r}s}) + {\rm h.c.}],
\label{EQ:2Dtsc}
\end{eqnarray}
which can be obtained, just as in $D=1$, by applying $-\pi/2$ spin rotation around the $z$ axis. This BdG Hamiltonian gives us $p_x+ip_y$ pairing between spin up electrons and $p_x-ip_y$ pairing between spin down electrons.%, and therefore the 1D helical Majorana edge state.
The $s$-wave superconductor on the square lattice has the Hamiltonian
\begin{eqnarray}
H_s =&-&\mu' \sum_{{\bf r'}s} f^\dagger_{{\bf r'}s} f_{{\bf r'}s} - t'\sum_{{\bf r'}s}\sum_{{\bf \hat{e}}={\bf \hat{x}},{\bf \hat{y}}} (f^\dagger_{{\bf r'}+{\bf \hat{e}} s} f_{{\bf r'}s}+ {\rm h.c.})\nonumber\\
&+&  \sum_{\bf r'}(\Delta' f^\dagger_{{\bf r'}\uparrow} f^\dagger_{{\bf r'}\downarrow} + {\rm h.c.}).
\end{eqnarray}
%is equivalent that of $D=1$ up to hopping in the $y$-direction.
The parameters we used were $|\Delta|=|\Delta'|$, $|t/\Delta|=|t'/\Delta'| = 3$, $\mu/|\Delta|=\mu'/|\Delta'|=-2$, and $|\delta t/t|=0.7$. We find that these parameters open up the edge state gap of $M/|\Delta| = 0.016$ at $\phi=\pi/2$; the fact that this deviates within an order of magnitude from the crossover value of $\tilde{h}$ indicated by Fig. \ref{FIG:2d} B. is because Eq. (\ref{EQ:ZeemanJJ2d}) assumes that of the edge state is linear and that the mass term is $k$-independent, neither of which is strictly true. For obtaining the result shown in Fig. \ref{FIG:2d} A. and B., we have set our superconductors on the lattices of 100 sites along the junction (the $y$-direction) and 30 sites perpendicular to the junction (the $x$-direction), and imposed a periodic boundary condition in the $y$-direction. For the numerical calculation in Fig. \ref{FIG:2d} D., we have 10 lattice sites along the $y$-direction (without periodic boundary condition) and 12 sites along the $x$-direction; in addition the perpendicular flux $\Phi$ modifies the hopping between the two superconductors:
\begin{equation}
H_1 = - \sum_{n=1}^{N_y}\sum_\sigma[(\delta t) e^{\frac{i(2n-N_y-1)\Phi}{4\Phi_0}}c^\dagger_\sigma (1,n)f_\sigma (m,n) + {\rm h.c.}].
\end{equation}
For the analytic calculation in Fig. \ref{FIG:2d} D., we used Eq. (\ref{EQ:ZeemanJJ2d}) with $M=0.02/\sin(4\pi/25)$, which is the value we infer from the numerical results for Fig. \ref{FIG:2d} B.

\section{Additional contributions from ordinary Josephson effect.}
%\begin{itemize}
%\item With SOC, there can be normal contribution to the Josephson coupling. This leads to a mixture of topological and ordinary Josephson effect.
%\end{itemize}
%We conclude that in the TRI TSC to the conventional SC junction in 2D, due to the Majorana edge state, we will have a Josephson coupling dominated by the second harmonics as long as the spin-orbit coupled hopping
It should be noticed that ordinary, bulk-to-bulk Josephson coupling between TSC and $s$-wave SC is possible through %in addition to the topological coupling due to Majorana edge states. At presence of
the spin-orbit coupled hopping across the junction, which would take the form
%\begin{equation}
$H_{SOC} = \sum_{n\sigma}\sum_{s=\pm 1} \lambda[\sigma c^\dagger_\sigma (1,n) f_{\bar{\sigma}} (m,n+s) + {\rm h.c.}]$ in $D=2$; this would lead to an ordinary contribution $I(\phi)\propto\sin\phi$.
%\end{equation}
%As long as its strength is small enough to satisfy $|\lambda/\delta t| \ll |\delta t/\Delta|^2$,
From perturbation theory, this conventional Josephson coupling will be much smaller than the second order Josephson coupling through the Majorana edge state in the limit of $|\lambda/\delta t| \ll |\delta t/\Delta|^2$; this is consistent with the singlet-triplet Josephson junction studied by Asano {\it et al.}\cite{asano2003} We have numerically studied the interplay between topological and regular Josephson effects %tuned by
with the spin-orbit coupling strength set at $\lambda/|\Delta| = 0.2$, which surely overestimates $\lambda$, with other parameters set the same as it was for Fig 3 (a) and we found the first harmonics to be less than 20\% of the second harmonics, even though $|\lambda/\delta t|$ and $|\delta t/\Delta|^2$ are of the same order of magnitude with our parameters. % outside the perturbative region.
%though our parameters were not in the $|\lambda/\delta t| \ll |\delta t/\Delta|^2$ limit.

\end{document}